\documentclass{pasj00}

\newcommand{\RXJ}{{RX~J1713.7$-$3946}}

\begin{document}
\SetRunningHead{Takahashi et al.}{Broad-band  X-ray spectrum of SNR \RXJ}
\Received{2007/06/16}
\Accepted{2007/08/03}

\title{Measuring the Broad-band X-Ray Spectrum from 400 eV to 40 keV
in the Southwest Part of the Supernova Remnant RX~J1713.7$-$3946}
\author{Tadayuki \textsc{Takahashi},\altaffilmark{1,2}
Takaaki \textsc{Tanaka},\altaffilmark{1,3}
Yasunobu \textsc{Uchiyama},\altaffilmark{1} \\
Junko S. \textsc{Hiraga},\altaffilmark{4}
Kazuhiro \textsc{Nakazawa},\altaffilmark{2}
Shin \textsc{Watanabe},\altaffilmark{1,2} 
Aya \textsc{Bamba},\altaffilmark{1} \\
John P. \textsc{Hughes},\altaffilmark{5}
Hideaki  \textsc{Katagiri},\altaffilmark{6}
Jun \textsc{Kataoka},\altaffilmark{7}
Motohide \textsc{Kokubun},\altaffilmark{1} \\
Katsuji \textsc{Koyama},\altaffilmark{8}
Koji \textsc{Mori},\altaffilmark{9}
Robert \textsc{Petre},\altaffilmark{10}\\
Hiromitsu  \textsc{Takahashi},\altaffilmark{6}
and
Yoko \textsc{Tsuboi}\altaffilmark{11}
}
\altaffiltext{1}{Department of High Energy Astrophysics,
  Institute of Space and Astronautical Science (ISAS), \\
  Japan Aerospace Exploration
  Agency (JAXA), 3-1-1 Yoshinodai, Sagamihara, 229-8510}
\altaffiltext{2}{Department of Physics, University of Tokyo, Hongo 7-3-1, Bunkyo, 113-0033}
\altaffiltext{3}{Stanford Linear Accelerator Center, 2575 Sand Hill Road, Menlo Park, CA 94025, USA}
\altaffiltext{4}{Cosmic Radiation Laboratory, RIKEN, 2-1 Hirosawa, Wako, Saitama 351-0198, Japan}
\altaffiltext{5}{Department of Physics and Astronomy, Rutgers University,\\ 136 Frelinghuysen Road, Piscataway, NJ 08854-8019, USA}
\altaffiltext{6}{Department of Physical Science, Hiroshima University, \\1-3-1 Kagamiyama, Higashi-Hiroshima, Hiroshima 739-8526, Japan}
\altaffiltext{7}{Department of Physics, Tokyo Institute of Technology, Ohokayama, Meguro, Tokyo, 152-8551,Japan}
\altaffiltext{8}{Department of Physics,  Kyoto University, Sakyo-ku, Kyoto 606-8502, Japan}
\altaffiltext{9}{Department of Applied Physics, 
University of Miyazaki, 1-1 Gakuen Kibana-dai Nishi 
Miyazaki, 889-2192}
\altaffiltext{10}{Astrophysics Science Division, NASA/Goddard Space Flight Center, Greenbelt, MD 20771, USA}
\altaffiltext{11}{Department of Physics, Chuo University, 1-13-27 Kasuga, Bunkyo-ku, Tokyo 112-8551, Japan}

\maketitle

\begin{abstract}
We report on results from Suzaku broadband X-ray observations of the southwest part of
the Galactic supernova remnant (SNR) \RXJ\ with an energy coverage of
0.4--40 keV. The X-ray spectrum, presumably of synchrotron origin,
is known to be completely lineless, making this SNR ideally suited for a detailed study
of the X-ray spectral shape formed through efficient particle acceleration at high speed shocks.
With a sensitive hard X-ray measurement from the HXD PIN  on board Suzaku,
we determine the hard X-ray spectrum in the 12--40 keV range
to be described by a power law with photon index $\Gamma = 3.2\pm 0.2$,
significantly steeper than the soft X-ray index of $\Gamma = 2.4\pm 0.05$ measured previously
with ASCA and other missions.
We find that a simple power law fails to describe the full spectral range of 0.4--40 keV
and instead a power-law
with an exponential cutoff with hard index $\Gamma = 1.50\pm 0.09$ and
high-energy cutoff $\epsilon_c = 1.2\pm 0.3\ \rm keV$ 
formally provides an excellent fit
over the full bandpass.
If we use the so-called SRCUT model, as an alternative model, 
it gives the best-fit rolloff energy of $\epsilon_{\rm roll} = 0.95\pm 0.04\ \rm
keV$. 
Together with the TeV $\gamma$-ray spectrum ranging from 0.3 to 100 TeV
obtained recently by HESS observations, our Suzaku observations of \RXJ\ provide
stringent constraints on the highest energy particles accelerated in a supernova shock.
\end{abstract}

\section{Introduction}

Over the past decade, the spectral and spatial properties of
non-thermal X-ray emission, presumably synchrotron
radiation by multi-TeV electrons, in young shell-type supernova remnants (SNRs)
have been extensively studied with X-ray satellites, especially with
ASCA and  Chandra,
providing observational support for the general picture that
Galactic cosmic rays are sustained by
strong shocks in SNRs
\citep{koyama95,slane99,gotthelf01,hwang02,uchi03,long03,bamba03,warren05}.
An excellent example comes from Chandra observations of
the Tycho's SNR, where the remnant is delineated by a thin
($1\arcsec$ or less) layer of X-ray synchrotron emission that exhibits spectral
steepening behind the shock \citep{cassam07}.
This strongly indicates that TeV-scale electrons are indeed accelerated
 at the outer blast wave.
The well developed theory of diffusive shock acceleration
(for reviews, see \cite{BE87,malkov01})
provides a basic framework for our understanding of cosmic-ray acceleration
in collisionless SNR shocks, though  some key ingredients  in the theory,
especially the generation of magnetohydrodynamic (MHD) waves and their interactions
with cosmic-ray particles \citep{bell01,bell04,DM07}, are yet to be understood.

X-ray observations of shell-type SNRs can in principle probe the ``microphysics"
of the shock acceleration process;
the development of MHD waves can be explored by measuring the energy of the
spectral cutoff that is regulated by the acceleration rate, which, in turn,
is determined by the shock velocity and the energy density of stochastic
magnetic fields relative to ordered fields.
In previous work, it has been presumed  that the nonthermal X-ray spectrum
is formed by a gradually steepening synchrotron spectrum extending from the radio
band, as first outlined by \citet{reynolds96}.
Experience shows that it is difficult
to establish the presence of  a cutoff (or roll off) in X-ray spectra
that cover only a decade or so in energy, like data typically obtained
by ASCA or Chandra.  Secure detection of the expected spectral steepening
requires a wider bandpass X-ray spectrum.
In contrast,
radio observations of SNRs sample the energy distribution of
accelerated electrons at considerably lower values (typically GeV energies)
where the spectral form is more nearly power-law like, and where the power-law
index in the framework of shock acceleration theory is closely related to
the shock compression ratio.

The SNR RX J1713.7$-$3946 is
a promising candidate to search  for a spectral
cutoff or steepening using measurements over as wide a
wavelength coverage as possible in the X-ray domain.
RX J1713.7$-$3946 (also known as G347.3$-$0.5)
exhibits the brightest nonthermal X-rays
among shell-type SNRs; the total nonthermal X-ray flux (2--10 keV)
integrated over the entire SNR amounts to
$\sim 5\times 10^{-10}\ \rm erg\ cm^{-2}\ s^{-1}$,
which is not exceeded by any other shell-type SNR.
This was the first SNR to be confirmed as a TeV $\gamma$-ray source
using ground-based Cherenkov telescopes by two groups,
CANGAROO \citep{muraishi00,enomoto02} and H.E.S.S. \citep{aha04,aha06a,aha07}.
Detections of
TeV $\gamma$-rays distributed over the entire remnant made with H.E.S.S.
with a 1--10 TeV flux of
$3.5\times 10^{-11}\ \rm erg\ cm^{-2}\ s^{-1}$,
make this object well suited for exploring cosmic-ray
acceleration in shell-type SNRs.
The spectacular  H.E.S.S. $\gamma$-ray image with spatial resolution of a
few arc-minutes appears generally quite similar to the X-ray image,
suggesting an intimate physical connection between the nonthermal X-rays
and TeV $\gamma$-rays.

Due to its  faintness in the radio band, the remnant eluded detection in
early radio surveys and was only discovered thanks to X-ray
observations during the ROSAT All-Sky Survey \citep{pfe96}.
ASCA revealed intense
synchrotron X-ray emission in the northwestern (NW) part of
the remnant \citep{koyama97}.
The spectrum here is best-fitted by a power law
of photon index $\Gamma =2.4\pm 0.1$ with interstellar absorption
$N_{\rm H} = (0.81\pm0.06)\times 10^{22}\ \rm cm^{-2}$, and is widely
presumed to be synchrotron in nature.
Diffuse synchrotron X-ray emission appears to be
distributed over the entire face of the SNR with similar index
($\Gamma \simeq 2.3\pm 0.2$) and without
measurable thermal X-ray emission \citep{slane99}.
Faint synchrotron radio emission was detected
in RX~J1713.7$-$3946, especially toward the western shell \citep{slane99,laz04}.
With Chandra observations, it was found that
the brightest part of the northwest X-ray shell
consists of a complex network of bright filaments and knots
embedded in more diffuse emission \citep{uchi03}.
The X-ray spectra of various  components
in the NW rim are similar to each other and well
fit by a power-law function with $\Gamma \simeq 2.1\mbox{--}2.5$ \citep{uchi03,laz04}.
Similar spatial and spectral characteristics were
found in the southwestern (SW) rim as well based on
\emph{XMM-Newton} data \citep{cassam04,hiraga05}.

A  nearby molecular cloud at $\sim 1$ kpc  is likely 
associated
with RX~J1713.7$-$3946,  based on high sensitivity CO observations
made with the NANTEN telescope \citep{fukui03,moriguchi05}.
The association between the molecular cloud at $\simeq 1$ kpc and the
X-ray remnant is 
suggested by their close morphological match as well as
by the detection of broad-line CO emission \citep{fukui03}.
This is further strengthened by a detailed comparison between
the X-ray and molecular emissions \citep{uchi05}.
The atomic and molecular hydrogen column density as a function
of distance along the line-of-sight to the SNR reaches a value
matching the X-ray absorbing column density at a distance of $\simeq 1\ \rm kpc$
\citep{koo04a}, thus providing
independent support for an association with the nearby (1 kpc) molecular cloud.
With this distance,  the guest star that appeared in A.D.~393
possibly corresponds to the supernova explosion
from which SNR RX~J1713.7$-$3946 originated \citep{wang97}.
This association would make the remnant 1600 yr old.
With this assumption,
an X-ray point source (without an optical/radio counterpart)
near the center of the remnant, 1WGA J1713.4$-$3949,
is thought to be an associated neutron star \citep{slane99,laz03},
indicating a core-collapse  supernova. It has been suggested
\citep{slane99,ellison01} that the supernova explosion occurred
within the wind-blown bubble
created by the massive progenitor star.
We take the distance to SNR RX~J1713.7$-$3946 to be 1 kpc in this paper.

Here we report on a wide band X-ray observation from 0.4 to 40 keV of
the SW part of  RX J1713.7$-$3946 made with the Suzaku satellite.
The high sensitivity in the 12--50 keV band offered by
the Hard X-ray Detector (HXD) on board Suzaku
together with the X-ray Imaging Spectrometer (XIS)
in the 0.4--12 keV band, allow us to measure for the first time
the detailed spectral shape of the nonthermal
X-ray emission from a supernova remnant.
Since this is the first clear example of detecting extended hard X-ray emission with the HXD,
we performed an in-depth analysis of the hard X-ray measurements.
In Section~\ref{sec:observation}, we describe the observations and methods of
obtaining the data sets. In Section~\ref{sec:analysis}, we explain the
analysis procedures to extract spectra and present spectral fitting
for each of the HXD and XIS in turn. We then proceed with the implementation of
HXD$+$XIS simultaneous modeling.
We discuss the broadband X-ray spectrum in terms of the theory of
diffusive shock acceleration in  Section~\ref{sec:discussion}.

\section{Observation and Data Reduction}
\label{sec:observation}

We observed the southwestern (SW) part of RX J1713.7$-$3946 with Suzaku
on 2005 September 26 (in the Performance Verification phase)
 for a total on-source duration of about 60~ks.
The brightest portion of the remnant is its northwest shell, which lies nearly
on the Galactic plane, where we were concerned that an unexpected
transient source could appear and contaminate the HXD field of view.
We therefore chose the second brightest part, i.e., the SW, to attempt a secure measurement of
hard X-ray emission with the HXD.
The X-ray Observatory Suzaku \citep{mitsuda06}, developed jointly by Japan and  the US, has
a scientific payload consisting of two kinds of co-aligned
instruments,  the XIS \citep{koyama_xis06} and HXD \citep{takahashi_hxd06,kokubun06}.
 The non-imaging HXD covers the hard X-ray bandpass of 10--600 keV with
 PIN silicon diodes (10--60 keV) and GSO scintillator (40--600 keV), both located
inside an active BGO shield.
 We here make use of  spectral data from the PIN diodes.
 The low background of the HXD PIN enables us to study high energy
emission from RX J1713.7$-$3946  up to several  tens of keV. Additionally,
a narrow FOV of $\sim 25\arcmin \times 25\arcmin$ (FWHM),
which is defined by  passive shields inserted
 in the BGO well-type collimator  above the PIN diodes,  is better suited to
 the study of extended emission in comparison with previous missions
having larger FOVs.
 The XIS consists of four X-ray imaging CCD cameras covering a FOV of
 $\timeform{17'.8} \times \timeform{17'.8}$;
 three are front-illuminated
 (FI: 0.4--12 keV), one is back side illuminated (BI: 0.2--12 keV), and all
have
 good energy resolution of $\sim 130$ eV at 6 keV.

\begin{figure}
\begin{center}
\FigureFile(8cm,8cm){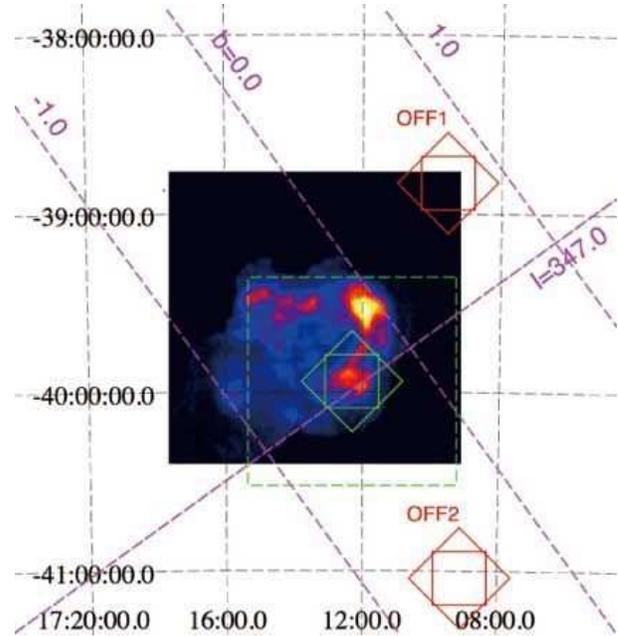}
\end{center}
\caption{Suzaku field of view for on-source (green) and off-source pointings (red),
overlaid on the ASCA GIS image (1--5 keV) taken from \citet{uchi05}.
North is up and east is to the left. Galactic coordinates are also shown.
The inner square corresponds to the FOV of the XIS observations and the outer diamond shows the rough shape of the
FOV (50\% effective area) of the PIN detectors
(see Fig.~\ref{fig:PINangular}).
The dashed square for the on-source demonstrates the full FOV of the PIN. }
\label{fig:rxj1713_obs_fov}
\end{figure}

The observation log for our observations is shown in Table \ref{tab:rxj1713_obs}.
In order to estimate possible contributions
from diffuse emission along  the Milky Way  in the hard 10--60 keV band,
 we observed two nearby regions in which no apparent hard X-ray point sources
exist (off-pointings, OFF1 and OFF2).
The three fields of view are depicted in Figure~\ref{fig:rxj1713_obs_fov},
overlaid on the ASCA 1--5 keV image \citep{uchi05}.
More recently, one year after the PV phase observations discussed here,
we have covered  almost the entire remnant  using 10 separate pointings
during the Suzaku AO-1 period. These results will be published elsewhere
(Tanaka et al., in preparation).

We used data products made with  version 1.2 of the pipeline processing.
For the XIS analysis, we retrieved ``cleaned event files'' which were screened
with standard event selections. We further screened the events
with the following criteria:
(1) cut-off rigidity larger than 8~GV and (2) elevation angle from the
Earth rim larger than $10^{\circ}$. For HXD data, ``uncleaned event files''
were screened with standard event screening criteria.
We excluded events during SAA passages and
Earth occultation, and also those with cut-off rigidity less than 8~GV.
The effective exposure times
after these screenings are shown in Table~\ref{tab:rxj1713_obs}.
In this paper,  data analysis was performed with  HEADAS 6.0.1.

\begin{table*}[htdp]
\caption{Summary of the {\it Suzaku} observations of RX~J1713.7$-$3946}
\begin{center}
\begin{tabular}{ccccc}
\hline
\hline
Pointing &  Obs. ID & Coord. (J2000) & Exposure & Date \\
&  & RA, DEC & XIS/HXD & \\
& &  &  [ks] & \\
\hline
\RXJ\ SW
& 100026010 & \timeform{17h12m17s.0}, \timeform{-39D56'11"} & 55/48 & 26/9/2005\\
\hline
OFF1 & 100026020 & \timeform{17h09m31s.9}, \timeform{-38D49'24"} & 28/24 & 25/9/2005\\
OFF2 & 100026030 & \timeform{17h09m05s.1}, \timeform{-41D02'07"} & 28/28 & 28/9/2005\\
\hline
\end{tabular}
\end{center}
\label{tab:rxj1713_obs}
\end{table*}%

\section{Analysis}
\label{sec:analysis}

\subsection{HXD Data Analysis}
\label{sec:hxd}

\begin{figure}
\begin{center}
\FigureFile(8cm,8cm){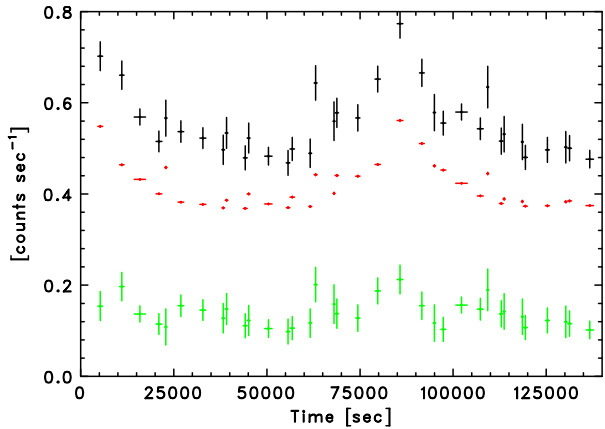}
\end{center}
\caption{HXD PIN lightcurves  in the 12--40~keV energy band.
The black, red, and green points show the raw,
background (NXB + CXB),
and background-subtracted data, respectively.
The NXB curve is constructed with the background model of \citet{watanabe07}.
The background-subtracted count rate is almost constant within statistical errors,
as expected for the extended object like this, except for high background periods. }
\label{fig:lightcurve}
\end{figure}

The HXD PIN achieves the lowest background level among all previous hard X-ray
instruments \citep{kokubun06}. Also a rather narrow FOV defined by a passive collimator
suppresses contamination from nearby discrete sources and diffuse emission such as
the Galactic ridge emission \citep{VM98} and the Cosmic X-ray background (CXB).
However, even for a bright object like this target,  the
 PIN spectrum is still exceeded by the background---time-variable instrumental
 background induced in one way or another by cosmic-rays and trapped charged
 particles in orbit.
 The HXD instrument team has developed an effective method \citep{watanabe07}
of modeling the time-dependent non-X-ray background (NXB)
by making use of the PIN upper discriminator (UD) signal that monitors passing charged particles through the silicon PIN diode. The background spectrum is
generated based on a database of NXB observations accumulated to date
during night- and day-earth observations.
Note that in our analysis the background spectral model was generated with 100 times
the actual number of background counts in order to minimize the photon noise
on the background.
The current NXB model is shown to be accurate within $\sim 5$\% \citep{mizuno06}.
In Figure~\ref{fig:lightcurve}, we show the light curve of the raw count rate in the 12--40 keV range as a function
of time for the on-source observation, the NXB count rate constructed with the model,
and the background-subtracted lightcurve.
It should be noted that the  background-subtracted lightcurve has to be
carefully examined in order to assure the validity of the background estimation.
Based on visual inspection, we discarded time regions when the total count rate was larger than
$0.6\ \rm counts\ s^{-1}$ since the background-subtracted lightcurve shows
an enhancement  in those time regions.
(When fitted with a power law,
the difference of photon index for fits with and without these time regions is
$\Delta \Gamma \simeq 0.2$.)
The source intensity after background subtraction
is about 50\% of the NXB level, implying that the systematic error on
the hard X-ray flux due to a miss-estimation of the background
should be about 10\%.

\begin{figure}
\begin{center}
\FigureFile(8cm,8cm){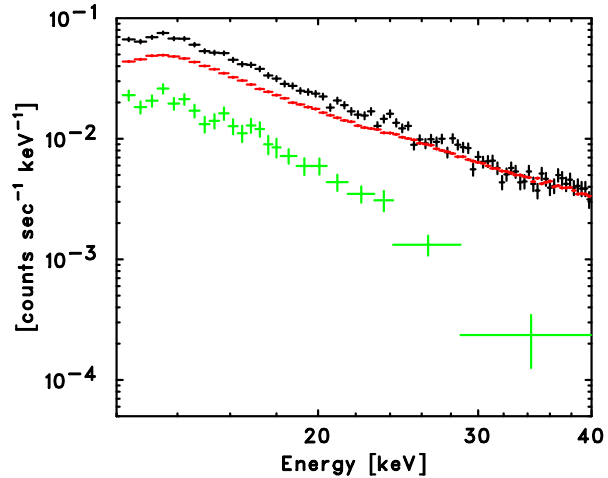}
\end{center}
\caption{Hard X-ray spectrum
of the southwest part of \RXJ\ recorded with the HXD PIN.
The data points in black
show the raw spectrum,  red points represent the background model
(NXB$+$CXB),
and green points are the background-subtracted spectrum.}
\label{fig:hxd_spec_bgdmodel}
\end{figure}

The HXD PIN spectra with and without background subtraction
are shown in Figure~\ref{fig:hxd_spec_bgdmodel}
together with the background model.
Since the NXB model does not include contributions from the CXB,
a simulated spectrum of the CXB was added to the NXB model.
Following \citet{gruber99}, who reanalyzed  HEAO-1 observations in the 1970's, 
our spectral model for the extragalactic background radiation was chosen as
\begin{equation}
\label{eq:CXB}
\frac{dN}{d\epsilon} = 7.9\, \epsilon_{\rm keV}^{-1.29} \exp \left( -\frac{\epsilon_{\rm keV}}{\epsilon_{\rm p}} \right)
{\rm ph}~{\rm s}^{-1}~{\rm keV}^{-1}~{\rm cm}^{-2}~{\rm str}^{-1},
\end{equation}
where $\epsilon_{\rm keV} = \epsilon/{1\ \rm keV}$ and
 $\epsilon_{\rm p} = 41.13$.
We estimated the expected CXB signal in the HXD-PIN spectrum
using the latest
response matrix for spatially uniform emission, {\tt ae\_hxd\_pinflat\_20060809.rsp}.
The contribution from the CXB flux is estimated to be $\sim 5$\% of the NXB,
comparable to the current systematic uncertainty of the NXB model itself.
We confidently detect hard X-rays from \RXJ\ up to $\sim 40$ keV; at higher
energies background uncertainty dominates the source flux.

\begin{figure}
\begin{center}
\FigureFile(8cm,8cm){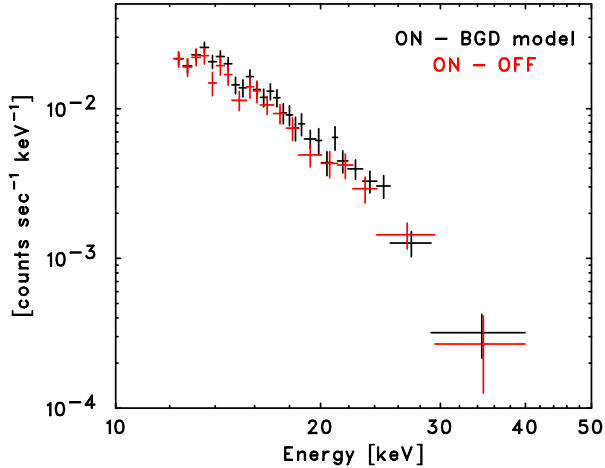}
\end{center}
\caption{Background-subtracted PIN spectra, using the background model (black)
 and off-source spectrum (red). }
\label{fig:hxd_spec}
\end{figure}

The PIN data from the two OFF-pointing observations were analyzed  in the same way as
those from the on-source pointing.
After subtraction of NXB and CXB,
we do not find any significant hard X-ray emission that exceeds the background
uncertainty. If we fit an NXB-subtracted spectrum of the OFF pointings
(OFF1 and 2 are added) with the CXB spectral form defined by Eq.~(\ref{eq:CXB})
but varying its normalization, we find a normalization factor of $1.7\pm 0.2$
(where the uncertainty includes only statistical errors).
This suggests the possible presence of diffuse emission from the Galactic plane \citep{VM98}
at a level of about 70\% of the CXB.  This is below the uncertainty in the NXB
model and will not significantly affect the
\RXJ\ spectrum. It should be noted that  the \emph{RXTE} PCA
spectrum of \RXJ\ in the 2--30 keV range
\citep{pannuti03} is likely contaminated by the Galactic ridge emission
because of its much larger FOV ($\sim 1^{\circ}$ FWHM).
Instead of the background model, we subtracted the OFF-pointing spectrum
from the on-source spectrum as shown in  Figure~\ref{fig:hxd_spec}.
We found that the two methods of background subtraction gave almost identical results
within the statistical error, though the OFF-subtracted spectrum seems
to be systematically lower than the model-subtracted spectrum.
This could indicate the level of the systematic error in the background model
or some contribution from the diffuse ridge component.

\begin{figure}
\begin{center}
\FigureFile(8cm,8cm){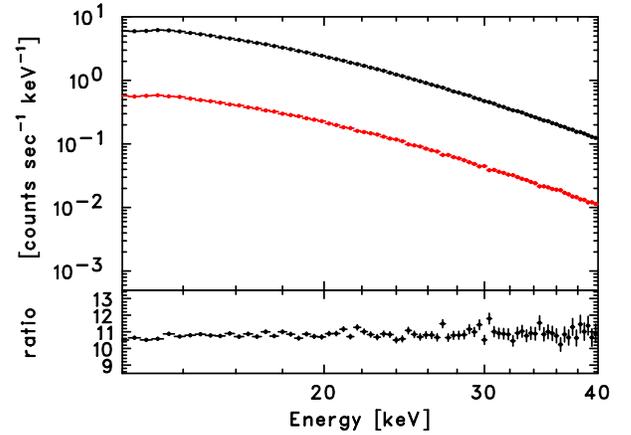}
\end{center}
\caption{Simulated PIN spectra for a point (red) and extended source (black).
The bottom panel shows the ratio between them. This demonstrates that the energy-dependence of
the PIN response for an extended source is same as that for a point source. }
\label{fig:ref_point_diffuse}
\end{figure}

\begin{figure}
\begin{center}
\FigureFile(8cm,8cm){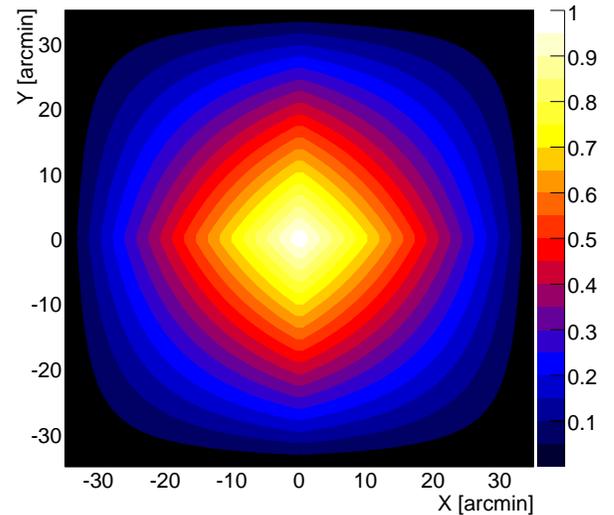}
\end{center}
\caption{Angular response of the PIN detector as defined by the built-in passive collimator.
Contours indicate the effective area as a function of incoming photon direction.
The angular response is independent of energy below 80 keV. }
\label{fig:PINangular}
\end{figure}

Before proceeding with the spectral fitting, we checked the detector response
against an {\it extended} source like \RXJ,
 since this is the first clear case that the Suzaku HXD
gives a spectral measurement for any source larger than the PIN collimator.
We performed Monte Carlo simulations with the code {\tt simHXD} \citep{terada05}
to study the PIN response for an extended source.
Hard X-rays from a $2^{\circ} \times 2^{\circ}$ sky area centered
on the optical axis were uniformly distributed over the entrance aperture of the HXD, and their interactions with the detector
system were tracked.
Figure~\ref{fig:ref_point_diffuse} shows the simulated
spectra of a diffuse source (red) and a point source at the XIS-nominal position (black).
The comparison demonstrates that  spatial extent does not affect
the observed spectral shape, consistent with expectations that
the built-in passive collimator ($50~\mu{\rm m}$ thick phosphor bronze sheets)
should almost completely absorb hard X-rays below 80 keV and, consequently,
that the angular response is not  energy-dependent.
In Figure~\ref{fig:PINangular} we show the angular response of
the PIN detector determined by the passive collimator.
The difference of the spectral shape is so small that it corresponds
to the difference of photon index of 0.01 when fitted with a power-law
model using the same response matrix.
We conclude that the extended nature of the source does not cause any spectral
steepening or flattening, at least for energies below 40~keV.
In the following, we therefore fit the PIN spectrum simply using the
point-source response
matrix at the XIS-nominal position,  {\tt ae\_hxd\_pinxinom\_20060814.rsp}.


We fitted the background-subtracted PIN spectrum with a simple power law model:
$dN/d\epsilon \propto \epsilon ^{-\Gamma}$.
The best-fit parameters are summarized in Table~2.
We present two cases, one using the NXB+CXB model as a background spectrum,
and the other using the OFF-pointings as a background template.
Both methods agree with each other.
Hereafter we only use spectral results with the NXB+CXB model.
The photon index obtained, $\Gamma = 3.2\pm 0.2$,  is much larger than that
for previous soft X-ray (below 10 keV) results, $\Gamma \simeq 2.4$ \citep{slane99}.
Given the current reproducibility of the NXB model of  $\sim 5$\%,
the systematic error due to uncertainties in the NXB modeling
is found to be $\Delta \Gamma \simeq 0.2$.
A spectral cutoff around 10~keV  seems necessary to account for the spectral transition
from the flat X-ray spectrum with $\Gamma =2.4\pm 0.05$ \citep{slane99}
  into the steep spectrum
with $\Gamma = 3.2\pm 0.2$ in the PIN domain.
Below we further confirm this important finding based on spectroscopy with the XIS.

\begin{table*}[htb]
\begin{center}
\caption{Power-law fitting to the HXD PIN spectrum\footnotemark[$*$]}
\begin{tabular}{ccccc}
\hline
\hline
Pointing & Background  & $\Gamma$ & Flux \footnotemark[$\dagger$] & $\chi_\nu^2 (\nu)$  \\
& & & (10--40~keV) & \\
\hline
\RXJ\ SW & model (NXB+CXB)
& $3.2\pm0.2$ & $ 2.5 \pm 0.1 $ & 1.15~(36) \\
\RXJ\ SW & OFF1+2
& $3.2\pm0.3$ & $ 2.3 \pm 0.2 $ & 0.92~(27)  \\
\hline
\multicolumn{5}{@{}l@{}}{\hbox to 0pt{\parbox{85mm}{\footnotesize
Notes.
\par\noindent
\footnotemark[$*$] Errors represent 90\% confidence.
\par\noindent
\footnotemark[$\dagger$] In units of mCrab.
}\hss}}
\end{tabular}
\end{center}
\label{tab:hxd_plfit}
\end{table*}

\subsection{XIS data Analysis}
\label{sec:xis}

\begin{figure*}
\begin{center}
\FigureFile(8cm,8cm){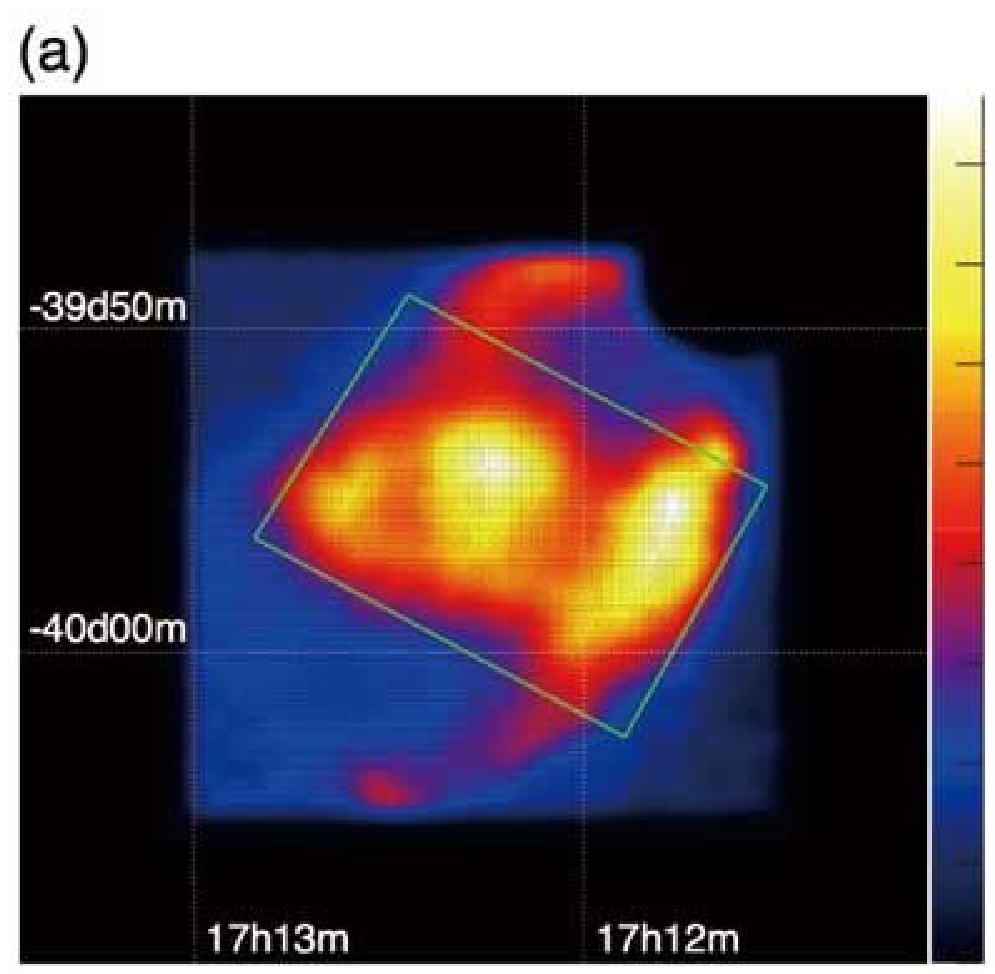}
\FigureFile(8cm,8cm){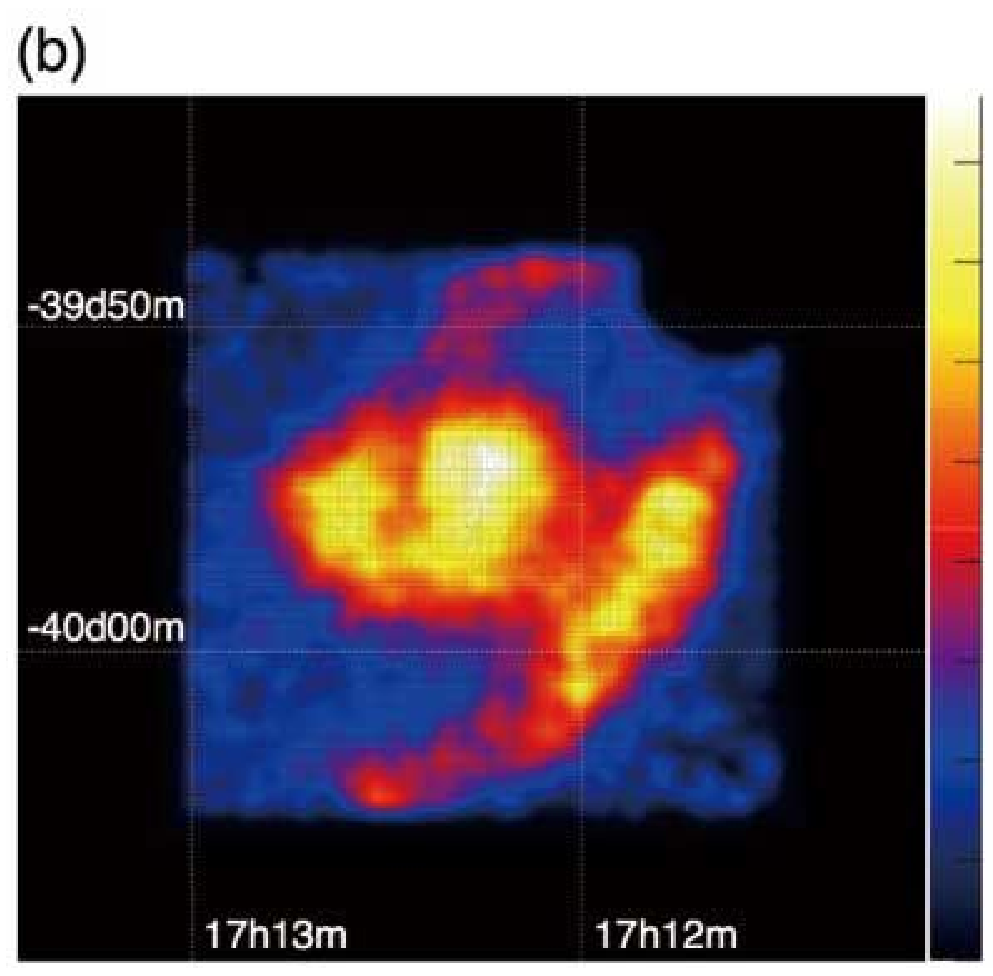}
\end{center}
\caption{The XIS  images of RX~J1713.7$-3946$ in (a) the soft 1--5~keV
and (b) the hard 5--10~keV bands. Three FI chips are combined. North is  up and
east is to the left. The images have dimension $17\arcmin \times 17\arcmin$.
The color scale indicates the surface brightness in a linear scale.
The green box shown in (a) is the region from which we extracted
the XIS spectrum.}
\label{fig:rxj1713_xis_image}
\end{figure*}

Figure \ref{fig:rxj1713_xis_image} shows the XIS images from our observation
in the soft (1--5~keV) and hard (5--10~keV) bands.
Both images were smoothed with a Gaussian kernel with $\sigma = 0^{\prime}.3$
and background-subtraction was carried out for each of them.
We utilized the Night Earth Background Database consisting of event data
accumulated while the satellite was viewing the night earth where the
non X-ray background becomes dominant.
After background subtraction, the vignetting effects of the X-ray mirrors were
corrected by means of the simulation program {\tt xissim} \citep{ishisaki06}.
It should be noted that the two XIS images in the soft and hard bands
are  similar to each other.  This indicates that spectral changes reported by
\citet{cassam04} across the SW part are not dramatic in our data set.

The XIS provides  higher quality spectra of the SNR \RXJ\ compared with previous missions
thanks to  its low background level, good energy resolution, excellent
energy response below 1 keV,  and large effective area \citep{koyama_xis06}.
We accumulated X-ray photons from the region shown in
Fig.~\ref{fig:rxj1713_xis_image}.
Since RX~J1713.7$-$3946 is situated near the Galactic plane, the background
spectrum needs to be accumulated from a nearby, source-free region.
We extracted the background spectra from the OFF observations.
In the fitting procedure, the standard RMF files (response matrices
of the XIS) version 2006-02-13 were used, whereas ARF files
(response of the XRT) were produced with the {\tt xissimarfgen}
software version 2006-10-26 \citep{ishisaki06}.
The energy range of 0.4--12 keV, to which the XIS FI chips are sensitive, is used in fitting, but
a range of 1.7--1.9~keV is excluded from the spectra
because of  large systematic uncertainties in the response matrices in this band.
Background uncertainty is most significant for the low energy ($<$0.6 keV) part of
the spectrum where the signal to background ratio drops.
We do not take account of this issue in detail, except to note that the
spectral fits reported here are not affected by this.
Finally, we co-added the three spectra, RMF files, and ARF files from the FI chips to produce a single data set for the XIS.

\begin{figure*}
\begin{center}
\FigureFile(8cm,8cm){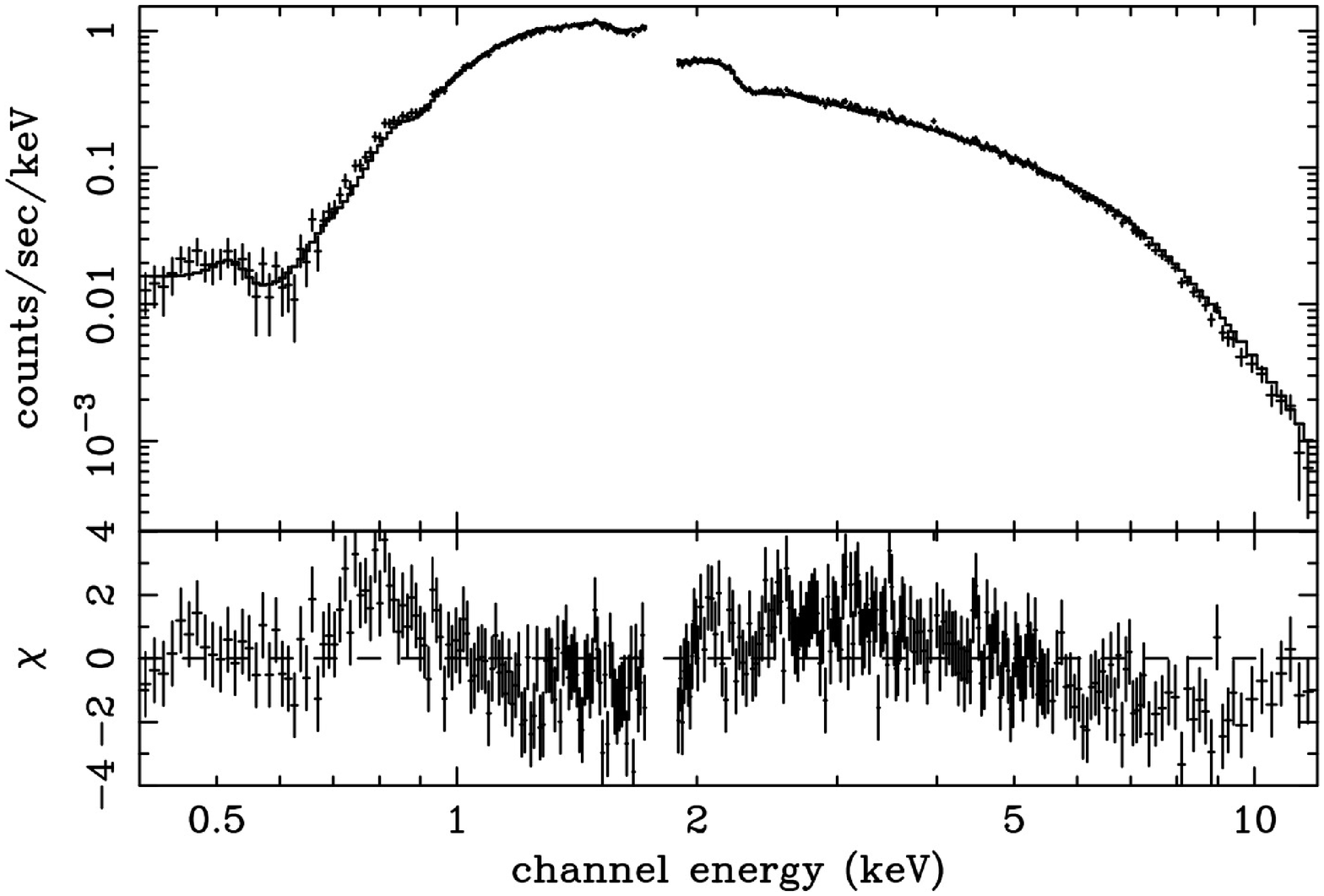}
\FigureFile(8cm,8cm){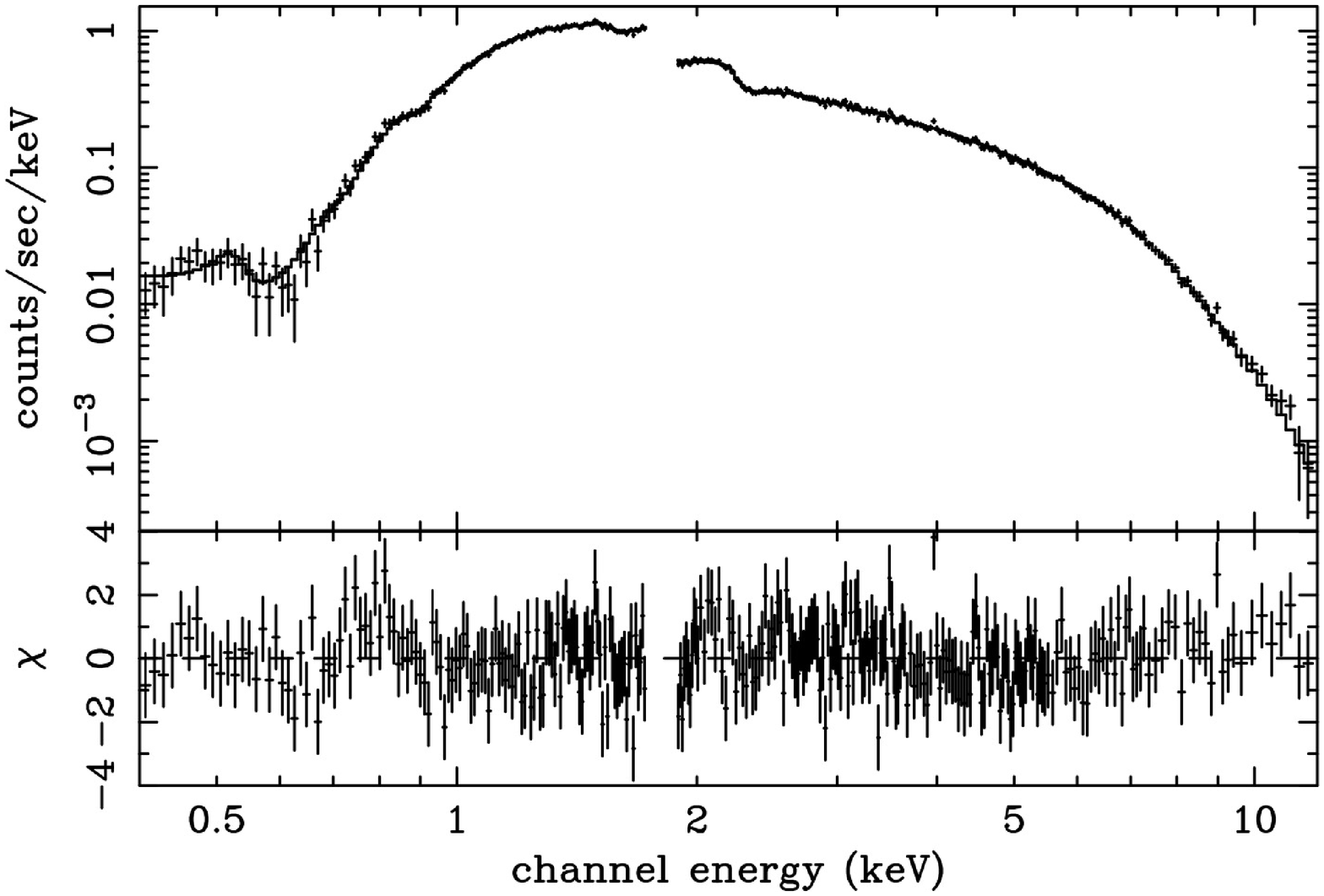}
\end{center}
\caption{(Left) The XIS spectrum in the 0.4--12 keV band
together with the best-fit power-law model. The lower
panel shows the residuals between the data and model.
(Right) Same as the left panel but with the best-fit cutoff power-law model. }
\label{fig:xis_fit}
\end{figure*}

Following previous studies, we first fitted the XIS spectrum in the
 0.4--12 keV band
with a simple power law attenuated by interstellar absorption.
We found a power law model fails to fit the data, giving a large reduced chi-squared of
$\chi_{\nu}^2=1.81$  for 302 degrees of freedom.
Though statistically unacceptable, the derived photon index of
$\Gamma = 2.39\pm 0.01$ and absorbing column density of
$N_{\rm H} = (0.87 \pm 0.01) \times 10^{22}\ {\rm cm}^{-2}$ are
in good agreement with the ASCA result in this region:
$\Gamma = 2.40\pm 0.05$ and
$N_{\rm H} = (0.79 \pm 0.50) \times 10^{22}\ {\rm cm}^{-2}$ \citep{slane99}.
In Figure~\ref{fig:xis_fit} (left panels) we show the XIS spectrum with the best-fit
power-law function and  residuals (data minus model).
There remains a correlated pattern to the residuals in the power-law fit.
This failure of the power-law fit argues that a more complicated
spectral model is required. Furthermore, the difference in fitted photon 
indices between the XIS and HXD PIN data says that the model needs
to roll-off or steepen with energy across the broad bandpass.

We therefore fit the XIS spectrum with three other models: a power-law
with two types of 
an exponential cutoff (cutoff power law), 
(1) $\epsilon^{-\Gamma} \exp\left[-(\epsilon/\epsilon_c)\right]$ and 
(2) $\epsilon^{-\Gamma} \exp\left[-\sqrt{\epsilon/\epsilon_c}\right]$, 
and (3) a broken power law. 
Unlike the pure power-law case,
these mathematical functions give acceptable fits (see Table~\ref{tab:xis_fit}).
Figure~\ref{fig:xis_fit} (right panels) shows the XIS spectrum with the best-fit 
$\epsilon^{-\Gamma} \exp\left[-(\epsilon/\epsilon_c)\right]$ function 
together with its residuals. The wavy residuals seen in the power-law fit clearly disappear
by introducing the cutoff.
The cutoff power-law fit gives a power law $\Gamma = 1.96\pm 0.05$ with
a cutoff at $\epsilon_c = 8.9^{+1.1}_{-0.9}\ \rm keV$, where the fit function has a form
$\epsilon^{-\Gamma} \exp\left[-(\epsilon/\epsilon_c)\right]$.
Also, we find 
a hard power law $\Gamma = 1.50\pm 0.09$ with
a lower cutoff at $\epsilon_c = 1.2^{+0.3}_{-0.2}\ \rm keV$ with the fit function of 
$\epsilon^{-\Gamma} \exp\left[-\sqrt{\epsilon/\epsilon_c}\right]$. 
The broken power-law model also reduces the correlated pattern in the residuals.
The best-fit function is composed of
two distinct power laws of $\Gamma_1 = 2.18\pm 0.03$ and
$\Gamma_2 = 2.53\pm 0.02$ with a break energy of $\epsilon_b = 3.13\pm 0.15$ keV.

We also performed spectral fits with the so-called SRCUT model
\citep{reynolds98} in the XSPEC package, 
which describes the synchrotron spectrum from
electrons following an exponentially-cutoff power law distribution in
energy.  In XSPEC, its spectral shapes for different parameters are tabulated 
based on numerical calculations 
convoluting a well-known synchrotron formula for a single electron with the energy 
distribution of electrons of the form $E^{-s}\exp( -E/E_0)$.  
The synchrotron spectrum has a power-law form at radio
frequencies (with photon index $\Gamma = \alpha +1$, where $\alpha$ is
nominally the spectral index in the radio band), with a rolloff energy
$\epsilon_{\rm roll}$ (nominally the energy where the exponential
cut-off reduces the flux by a factor of 4 below the extrapolation of
the radio power-law spectrum).  We fixed the index to a typical
value for SNRs of $\alpha = 0.5$.  A good fit was obtained with a
best-fit rolloff energy of $\epsilon_{\rm roll} = 0.95\pm 0.04\ \rm
keV$.
The roll-off energy obtained in this way is similar to the cutoff energy 
$\epsilon_c = 1.2$ keV obtained 
for a spectral form $\epsilon^{-\Gamma} \exp\left[-\sqrt{\epsilon/\epsilon_c}\right]$.
This is reasonable because the SRCUT model has an approximate form 
$\epsilon^{-\Gamma} \exp\left[-\sqrt{\epsilon/\epsilon_{\rm roll}}\right]$ 
in a narrow bandpass around the cutoff region. 

\begin{table*}[htdp]
\caption{Model fitting to the XIS spectrum (0.4--12 keV)\footnotemark[$*$]}
\begin{center}
\small
\begin{tabular}{cccccc}
\hline
\hline
Function $\frac{dN}{d\epsilon}$ &$N_{\rm H}$  &  $\Gamma$ & $\epsilon_c/\epsilon_b/\epsilon_{\rm roll}$
& $F_{1-10~{\rm keV}}$\footnotemark[$\dagger$] & ${\chi_{\nu}}^2~(\nu)$\\
  & $(10^{22}~{\rm cm}^{-2})$ & & (keV) & $(10^{-11}{\rm erg}~{\rm s}^{-1}~{\rm cm}^{-2})$ & \\
\hline
power law ($\epsilon^{-\Gamma}$)
& $0.87\pm0.01$ & $2.39\pm0.01$ & ---  & 6.07 $\pm$ 0.09 & 1.81~(302) \\
$\epsilon^{-\Gamma} \exp\left[-(\epsilon/\epsilon_c)\right]$
 & $0.77 \pm 0.01$ & $1.96\pm0.05$ & $8.9^{+1.1}_{-0.9}$& $5.67^{+0.10}_{-0.05}$ & 1.03~(301) \\
$\epsilon^{-\Gamma} \exp\left[-\sqrt{\epsilon/\epsilon_c}\right]$
 & $0.74 \pm 0.02$ & $1.50\pm0.09$ & $1.2^{+0.3}_{-0.2}$& $5.61\pm 0.40$ & 1.00~(301) \\
broken power law & $0.79\pm 0.01$  & $2.18\pm 0.03$/$2.53\pm 0.02$ & $3.13\pm 0.15$
& $5.75\pm 0.17$ & 1.00~(300) \\
srcut  & $0.77\pm 0.01$  & $1.5_{\ \rm fixed}$ & $0.95\pm 0.04$
& $5.71\pm 0.02$ & 1.05~(302) \\
\hline
\multicolumn{6}{@{}l@{}}{\hbox to 0pt{\parbox{85mm}{\footnotesize
Notes.
\par\noindent
\footnotemark[$*$] Errors quoted at 90\% confidence.
\par\noindent
\footnotemark[$\dagger$] Corrected for absorption.
}\hss}}
\end{tabular}
\end{center}
\label{tab:xis_fit}
\end{table*}%

\subsection{Suzaku Wide Band Spectrum}
\label{sec:wideband}

We now combine the HXD PIN spectrum with the XIS spectrum
to fully exploit the broadband capability of Suzaku.
Since the source is extended and the HXD PIN observes a
much wider field than the XIS,
the relative normalization factor derived for a point source,
$\kappa = 1.13\pm 0.01$ \citep{ishida06},  is not applicable in our case.
We note that the relative normalization factor $\kappa$ is defined 
here as the flux density
at 12 keV determined by the PIN relative to  that determined by the XIS.
Therefore, it is necessary to calculate the proper scaling factor between the
XIS and the HXD for a diffuse emission source.
We determined the normalization factor $\kappa$,
which takes into account the larger FOV of the PIN,
by convolving the emission distribution of the ASCA image with
the angular response of the HXD-PIN defined by the passive collimator.
It is assumed that the spatial distribution of
X-ray surface brightness is independent of energy
in the ASCA-Suzaku band over the entire remnant.
We determined the normalization factor to be
$\kappa = 6.6$  using  the ASCA 5--10 keV image, with a
possible systematic error of $\sim 10\mbox{--}20\%$.
(The value of $\kappa$ does not depend sensitively on the choice of
ASCA bandpass. If instead we use the ASCA 1--5 keV image a slightly
different value, $\kappa = 6.5$, is obtained.)

\begin{figure*}
\begin{center}
\FigureFile(16cm,16cm){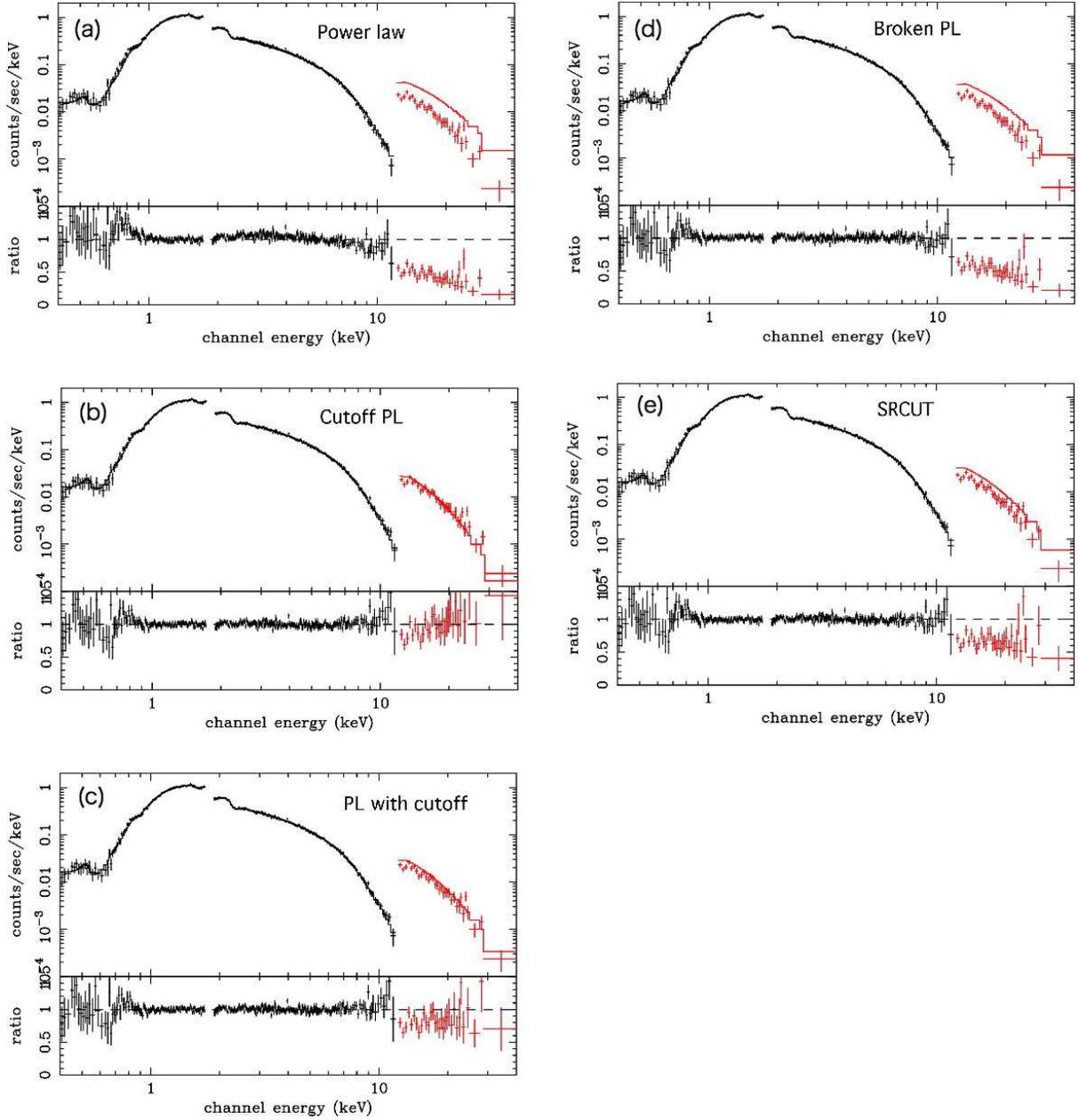}
\end{center}
\caption{XIS$+$PIN spectra in the energy range 0.4--40 keV are shown 
with the best-fit models obtained by fitting the XIS spectrum alone: 
(a) power law, (b) $\epsilon^{-\Gamma} \exp\left[-(\epsilon/\epsilon_c)\right]$
(c) $\epsilon^{-\Gamma} \exp\left[-\sqrt{\epsilon/\epsilon_c}\right]$
(d) broken power-law, and  (e)  SRCUT. 
The lower panels show the ratio between the date points and the model values.}
\label{fig:wide}
\end{figure*}

In Figure~\ref{fig:wide}, we show the XIS$+$PIN spectra together with
the power law, cutoff power laws, broken power law, and SRCUT models.
The model parameters are the best-fit values determined by fitting the
XIS spectrum alone (see Table~\ref{tab:xis_fit}), and the best-fit
models are extrapolated to the PIN bandpass after multiplication by
the factor $\kappa = 6.6$.  The power-law model clearly conflicts with the PIN
data as does the broken power law model, which was a good description
of the XIS data alone.  Although the shape of the SRCUT model in the
PIN band (which can be characterized as a power law with a photon
index of $\Gamma \simeq 3.3$) is a good match to the data, the level
of hard X-rays this model produces is about 40\% too high, somewhat
more than the estimated systematic error in the normalization factor
$\kappa$ of $\sim 20\%$. If we change the exponent, $\alpha$, of 
the SRCUT model by $\pm 0.1$, this result still holds. 
 Although the cutoff power-law function of 
 $\epsilon^{-\Gamma} \exp\left[-(\epsilon/\epsilon_c)\right]$
satisfactorily reproduces the level of the observed PIN flux, the
model flux in this case (approximately a power law with $\Gamma \simeq
3.7$ in the PIN band) falls more steeply than the data. 
A slower exponential form of 
$\epsilon^{-\Gamma} \exp\left[-\sqrt{\epsilon/\epsilon_c}\right]$, 
which gives $\Gamma \simeq 3.5$ in the PIN band for its best-fit model, 
seems to yield the best match to the PIN data both in the absolute flux and 
spectral shape.
We conclude that the evidence
for a curvature in the broad band Suzaku spectrum of \RXJ\ is secure.
However, it is premature at this time to argue that, for example, the
cutoff power-law models are a better description of the emission
conditions in the SNR than the SRCUT model based on the XIS$+$PIN
joint spectral analysis, simply because the two data sets do not come
from the same spatial regions.  Modest spatial variations in the
emitting conditions could produce flux or spectral differences
similar to those seen in the PIN band for the cutoff power-law and
SRCUT models.  Furthermore, even spatial/spectral variations in the
synchrotron emission within the joint XIS$+$PIN field of view (as
seen by \citet{cassam04}) could account for the modest differences in the
cutoff power-law and  SRCUT models in the PIN band.

\subsection{Upper Limits on Thermal Emission}
\label{sec:thermal}

One of the most remarkable characteristics of RX~J1713.7$-$3946 is the complete lack
of a thermal X-ray emission component as expected from the hot shocked gas in
the  SNR.
Our XIS spectrum, similar to previous studies \citep{slane99,cassam04},
shows no signs of thermal emission.
In the following we set an upper limit on the density of  shocked hot gas from
the upper limit on the thermal emission measure.

To derive the allowable level of thermal emission underneath the synchrotron
spectrum, we added a thermal component to the cutoff power law model
obtained by fitting the XIS spectrum (see Table~\ref{tab:xis_fit} for the parameters).
We used the APEC model for the thermal emission model and assumed solar composition.
Then, we determined 3$\sigma$ upper limits on the emission measure,
$\mathrm{EM} = \int n_{\mathrm e} n_{\mathrm H} dV/4 \pi D^2$,
as a function of electron temperature
$kT_\mathrm{e}$ over the range 0.05~keV to 1~keV. Above 1 keV, the upper limit
has only a weak dependence of temperature \citep{slane99}.
Here, $n_\mathrm{e}$, $n_\mathrm{H}$, $V$, and $D$ are the number density of 
electrons and hydrogen, the volume of the hot gas, and the distance to the remnant, respectively.

\begin{figure}
\begin{center}
\FigureFile(8cm,8cm){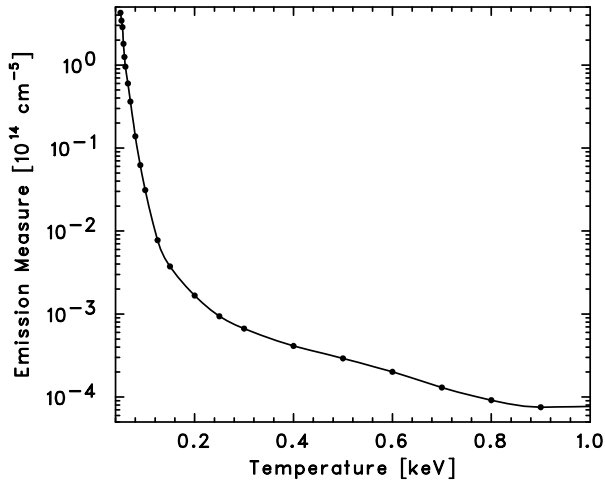}
\end{center}
\caption{The 3$\sigma$ upper limit on thermal emission in \RXJ\ expressed
as emission measure
(${\rm EM} = \int n_{\mathrm e} n_{\mathrm H} dV/4 \pi D^2$)
vs.\ electron temperature $kT_{\rm e}$.}
\label{fig:temp_vs_EM}
\end{figure}

Figure~\ref{fig:temp_vs_EM} shows the 3$\sigma$ upper limits as
a function of temperature. The upper limits grow quite large for
$kT_{\mathrm e} \lesssim 0.2~\mathrm{keV}$.
The emission measure can be as large as $10^{14}~\mathrm{cm}^{-5}$ if we
assume a temperature as low as $\sim 0.05$~keV. From the emission
measure, we can estimate the number density of shocked gas.
By assuming $n_{\mathrm e} = 1.2 n_{\mathrm H}$, uniform
distribution of gas inside the remnant, and that the spectral region covers 8\% of the
whole volume of the remnant (a sphere of $30^{\prime}$ radius),
we relate the gas density ($n \equiv n_{\mathrm H}$) to the emission measure as
\begin{eqnarray}
n =1 \ \left( \frac{\mathrm{EM}}{10^{14}~\mathrm{cm}^{-5}} \right)^{1/2}
\left( \frac{D}{1~\mathrm{kpc}} \right)^{-1/2}~\mathrm{cm}^{-3}.
\end{eqnarray}
For a distance of $D=1\ \rm kpc$,
the upper limit on the gas density  ranges from
$9 \times 10^{-3}~\mathrm{cm}^{-3}$ to $2~\mathrm{cm}^{-3}$ when we
vary the temperature from 1~keV to 0.05~keV.
This is broadly consistent with previous estimates based on ASCA \citep{slane99} and XMM-Newton \citep{cassam04}.

\section{Discussion and Conclusions}
\label{sec:discussion}

As established in previous studies 
(\cite{koyama97,slane99,uchi03,laz04,cassam04,hiraga05}),
the X-ray spectrum extracted from any portion of RX J1713.7$-$3946
shows a featureless continuum, for which an absorbed power law offers a reasonable fit.
The ASCA spectrum of the SW part was characterized by a photon index
$\Gamma = 2.40\pm 0.05$ \citep{slane99}.

The Suzaku broadband spectrum for the SW part of \RXJ\ reported here clearly
indicates the presence of a spectral cutoff  as it shows
a spectral transition from a relatively
flat power law ($\Gamma \sim 2.3$) in the soft X-rays
to a steep power law ($\Gamma \sim 3.3$) at harder photon energies.
In fact, the nonthermal featureless X-ray spectrum of \RXJ, as well as other
nonthermal remnants such as SN~1006 \citep{koyama95} and Vela Jr.\ \citep{slane01},
are widely thought to arise from the gradually steepening
high-energy part of the synchrotron radiation
spectrum produced by TeV-scale electrons, which are accelerated
through  the mechanism of diffusive shock acceleration \citep{reynolds98}.
The cutoff in the Suzaku spectrum provides  firm observational support for
this widely-accepted presumption.
We found that the cutoff energy, presumably set by 
synchrotron cooling, appears different from one spectral model to another, 
 in a range of $\epsilon_{\rm cutoff} \sim 1\mbox{--}9$ keV. 
The higher cutoff, $\epsilon_c \sim 9$ keV, is based on 
the power law with $\exp(-\epsilon )$-type cutoff 
model, while  the lower bound is derived with 
$\exp\left(-\sqrt{\epsilon}\right)$-type cutoff. 
Given this fact, we need to be careful in discussing the physical meanings of 
the position of the spectral cutoff. 
We briefly explore the implications of the Suzaku broadband spectrum 
in a separate paper \citep{uchi07}.  More detailed discussions on this issue 
will be given elsewhere (T. Tanaka et al.\ in preparation).

If the magnetic field exceeds $10~\mu \rm G$ in the shell (e.g., see \citet{BV06}), 
the TeV $\gamma$-rays detected from the remnant should be explained by 
protons producing $\pi^0$-decay $\gamma$-rays, rather than by electrons 
emitting $\gamma$-rays via inverse Compton scattering \citep{aha06a}.
The hadronic model requires the total energy content of protons to amount to
\begin{eqnarray}
W_p \approx 10^{50} \left( \frac{D}{1~\mathrm{kpc}} \right)^2
\left( \frac{n}{1~\mathrm{cm}^{-3}} \right)^{-1} \ \mathrm{erg},
\end{eqnarray}
in order to explain the observed TeV $\gamma$-ray flux \citep{aha06a}.
Therefore, a matter density of $\sim 0.2~\mathrm{cm}^{-3}$ is needed, assuming the
typical kinetic  energy released by a supernova of $10^{51}$~erg and a conversion
efficiency to high energy protons of 50\%.
To reconcile this density value with the upper limit on thermal X-ray
emission (\S\ref{sec:thermal}),
the electron temperature must be of order 0.1$~$keV (or lower),
otherwise thermal emission ($\propto n^2$) would have been detected
with the Suzaku XIS.
The low-temperature of thermal electrons may indicate
that efficient particle acceleration drains the energy of the shock,  thus
suppressing gas heating at the shock \citep{ellison01}.

Strong (high Mach number) shocks in young shell-type SNRs do accelerate
high-energy particles with remarkably high efficiency, thus
providing an ideal laboratory to study
high-energy particle acceleration in the Universe.
Suzaku has excellent capability for determining the spectral form
of X-ray synchrotron emission thanks to its broadband spectral coverage
from 0.4 to 40 keV.
The observed cutoff energy in the synchrotron X-ray spectrum
suggests  that shock acceleration proceeds at nearly the maximum possible rate.
Combined with the broad  TeV $\gamma$-ray spectrum (in the 0.3--100 TeV band)
measured recently with HESS, our Suzaku observations of \RXJ\ provide
stringent constraints on shock-acceleration of the highest energy particles
in our Galaxy.

\end{document}